\documentclass[sn-mathphys]{sn-jnl}

\usepackage{algorithm}
\usepackage{algorithmicx}
\usepackage{algpseudocode}
\usepackage{tabularx}
\usepackage{longtable}
 \usepackage{graphicx}
 \usepackage{lscape}

\jyear{2021}%

\theoremstyle{thmstyleone}%
\theoremstyle{thmstyletwo}%

\theoremstyle{thmstylethree}%

\begin{document}
	
	\title[ ]{Human Activity Behavioural Pattern Recognition in Smart Home with Long-hour Data Collection}
	
	
	\author*[1]{\fnm{Ranjit} \sur{Kolkar}}\email{ranjit.kolkar@gmail.com}
	
	\author[1]{\fnm{Geetha} \sur{V}}\email{geethav@gmail.com}
	\equalcont{These authors contributed equally to this work.}

	\affil[1]{\orgdiv{Department of Information Technology}, \orgname{National Institute of Technology Karnataka}, \orgaddress{\street{Srinivasnagar}, \city{Surathkal}, \postcode{575025}, \state{Karnataka}, \country{India}}}

\abstract{The research on human activity recognition has provided novel solutions to many applications like healthcare, sports, and user profiling. Considering the complex nature of human activities, it is still challenging even after effective and efficient sensors are available. The existing works on human activity recognition using smartphone sensors focus on recognizing basic human activities like sitting, sleeping, standing, stair up and down and running. However, more than these basic activities is needed to analyze human behavioural pattern. The proposed framework recognizes basic human activities using deep learning models. Also, ambient sensors like PIR, pressure sensors, and smartphone-based sensors like accelerometers and gyroscopes are combined to make it hybrid-sensor-based human activity recognition. The hybrid approach helped derive more activities than the basic ones, which also helped derive human activity patterns or user profiling. User profiling provides sufficient information to identify daily living activity patterns and predict whether any anomaly exists. The framework provides the base for applications such as elderly monitoring when they are alone at home. The GRU model's accuracy 95\% is observed to recognize the basic activities. Finally, Human activity patterns over time are recognized based on the duration and frequency of the activities. It is observed that human activity pattern, like, morning walking duration, varies depending on the day of the week. }

	
	
	
	\maketitle
	
	\section{Introduction}
	\label{sec:introduction}
	
The population of the elderly is growing much faster, according to the survey \cite{bib1}. Life expectancy is increasing from 71 years to 77 in the next three decades. Many Asian countries, including India, have a larger population of aged people, even though the rate at which it is increasing will be more than 20\% by 2050. This has brought the potential to develop long-term care, primitive health assistance, and long-distance monitoring, which improves the quality of life\cite{bib2}. The advancement in sensor hardware with the increased ability to perform better has opened many research areas in assisting people. The smart-home system is the domain that addresses all these problems to improve people's quality of life\cite{bib3}. Smart Home is equipped with a smart application that can sense the environment and can be controlled by the owner. A Human Activity Recognition(HAR) system studies the identification of certain activities performed in a context with the help of wearable, vision, and ambient sensors\cite{bib4}. HAR systems are deployed to give personalised health assistance in a smart home, along with effective and efficient resource utilisation.

Human Activity Recognition(HAR) correctly recognises human activities or actions in their work context. In recent years, sensors have attained significant noteworthiness in research and application. With the increased capabilities of the sensors, the applicability of these sensors has increased in many real-time applications. HAR has many applications in sport and physical activities\cite{bib5}\cite{bib6}\cite{bib7}, interactive games\cite{bib9}, surveillance\cite{bib8}, human-computer interaction\cite{bib10}, rehabilitation, ergonomics, monitoring older people, patients etc. In sports, the activity recognition of a sportsman can be analysed and improve performance. In surveillance\cite{bib11}, HAR is used in smart campuses, smart homes, and smart health to detect the abnormal behaviour of humans, and suitable actions can be taken. 

There are already existing datasets for human activity recognition using smartphone-based data with accelerometers and gyroscope sensors\cite{bib12}. An accelerometer is a sensor device that monitors the acceleration of any person or object in its immediate rest frame. Several accelerometer sensor-based devices include wearable devices, smartphones, and other electrical equipment to recognise human activities and their activity patterns. However, these sensors provide the data to recognise only basic human activities like walking, sleeping, running, and sitting. This work focuses on finding the patterns of human activities over a long period (Ex. daily, weekly). With the existing dataset\cite{bib13}, which uses only smartphones, it is challenging to have continuous daily living activities. Hence the work focuses on creating a dataset that records human activity over a long period in a home environment. We have used the existing mobile phone sensors to avoid the burden of wearing specific wearable devices and or other sensors. The data recording is done in three phases. First, the six basic activities are recognised using the deep-learning models on the WISDM dataset. Second, the trained model is deployed in the mobile application to recognise the basic activities, and the ambient sensor data is recorded simultaneously. In the third phase, the basic activities and the ambient sensor data are combined to derive more activities like watching TV in Hall.

With smartphone-based sensors, the activities that can be recognised are limited to walking, stairs up, stairs down, sleeping, and falling \cite{kolkar2021issues}. The limitation is because of the placement of the mobile phone inside the pocket. With mobile phone-based sensors, the activities done using hand movements cannot be recognised. However, using the ambient sensors attached to the bed, chair, TV, and location-detecting sensors, it is possible to derive more activities like using the bathroom, using the kitchen, watching TV, sleeping, and sitting on a chair. Table \ref{table:Activities_List} briefs the activities from the mobile sensors and activities derived from the ambient sensors. So, an Android-based application is developed to collect the data for long-hour. The user needs to configure it by specifying the basic parameters of sensors and duration at the beginning. To develop the Android application, we used the Android software development toolkit and android studio as an integrated development environment to design the user interface and develop the application. 

The proposed dataset can be used to analyse a person's activity patterns, which can be applied to various applications such as personality profiling, anomaly detection, and habit recording. The time-series data's feature set is used for deep learning-based classification. Accurate real-time human activity recognition is designed using deep learning models such as LSTM and GRU. Performance parameters such as accuracy, F-1 score, recall, and precision are used to evaluate the learning model. The real-time recognition model is then deployed on the smartphone and given to the user for data recording for a long hour. The activities are labelled automatically using the trained model deployed in phase 2, with one activity per minute being sampled. The authors then manually verify the label in the smartphone application to correct false labelling done by the mobile application.

The work focuses on the human activity recognition system using smartphone-based sensors. Many datasets \cite{bib20}  \cite{bib21}  \cite{bib22}  \cite{bib23} are available on which the experiment is being done for HAR systems. However, these data sets are lab-controlled environments and only recognise the basic activities. So, the long-hour dataset is created to detect human activity patterns over a period. This work has the following contributions.  

	\begin{itemize}
	
	\item  Designed the data fusion on the raw data of the existing WISDM dataset to improve the performance of the HAR system. Improved accuracy is observed in the recognition of activities with data fusion. 
	\item  The android-based application is developed and deployed with a trained model and automatically labels the new dataset for basic human activities. 
	\item  Ambient sensors output is synchronized with the long-hour data to derive more activities. The effective configuration and placement of PIR sensors aided in detecting the room in which the person resides. The simple relay switches On/Off detection circuit also assisted in deriving more activities than the basic ones.
	\item  Single person occupancy detection algorithm is developed to detect the person inside the room and resolve the occupancy overlaps between the rooms.
	\item The priority-based Two-Minute segmentation of data that gives a label every 2 minutes is designed. This labelled data is mapped to derived activities. Priority-based averaging is used based on person-centric activities, which helps reduce the data size and applies very well in the application context.  
\end{itemize}

The motivation behind this work is to understand human behaviour and activity patterns for applications such as healthcare, user profiling, and remote monitoring. By accurately recognizing and analyzing human activities using a hybrid-sensor-based approach, combining smartphone-based and ambient sensors, personalized interventions can be developed, training regimens can be optimized, and overall well-being can be enhanced. This framework aims to improve the accuracy and richness of human activity recognition by leveraging advancements in sensor technologies and deep learning models. The derived activity patterns enable the identification of daily living patterns, prediction of anomalies, and facilitate targeted interventions in areas such as elderly monitoring, performance optimization, and healthcare management. 

This work represents an ongoing research effort, laying the foundation for a comprehensive analysis of human activity patterns. Building upon this work, the goal is to thoroughly explore activity patterns based on the activity duration, frequency and sequences. We are experimenting with adaptive granularity based on grouping and temporal intervals to provide a generalized pattern analysis. By leveraging the insights gained from this ongoing work, a comprehensive understanding of human behaviour will be achieved, leading to advancements in personalized recommendations, anomaly detection, and overall quality of life improvements.

The paper is further presented in many sections; Section 2 lists the related work. Section 3 focuses on how the proposed methodology for the long-hour dataset is created in phases. The implementation details and results are discussed in Section 4. Section 5 summarises human behaviour activity pattern recognition, Section 6 discusses the limitations of the proposed work and Section 7 presents the conclusion and future work.

\section{Related Work}

Solving the daily safety and nursing problems of the elderly is an essential purpose and application field of activity recognition research. The proposed a hierarchical activity recognition system\cite{bib14} to identify abnormal activities (such as chest pain, headache, vomiting, syncope, etc.) in the lives of elderly people living alone. An automatic periodic monitoring system to ensure the safety of the elderly and improve the quality of life of the elderly people living alone, as well as reduce the cost of health care. The authors\cite{bib15} focused on recognising the daily activities recognition of the elderly to help doctors or family members to understand the regularity of their activities and the amount of exercise and improve the accuracy of doctors' health evaluation to formulate more appropriate rehabilitation strategies. The authors\cite{bib16} designed a wearable fall recognition and alarm system using a 3-axis accelerometer and a gyroscope to monitor the activities of the elderly and timely use the mobile phone to alarm. The system can support quickly carrying out the rescue work after the fall. The literature work shows that using simple sensors like accelerometers and gyroscopes, elderly monitoring is possible. However, challenges like limited activity recognition, short-duration data, and lack of activity patterns were observed.

Much literature has been about human activity recognition using wearable-based sensors with Artificial Intelligence. The authors\cite{bib13} proposed a deep convolutional neural network (CNN) approach for recognising human activities from wearable sensor data. The proposed model was trained and tested on the WISDM dataset and achieved an accuracy of 95.4\%. This paper\cite{bib14} compared various feature extraction techniques and machine learning algorithms for human activity recognition using wearable sensors and smartphones. The authors found that support vector machines (SVMs) with a polynomial kernel outperformed other algorithms in accuracy. This paper comprehensively reviews recent advances in human activity recognition using smartphones and wearable sensors. The authors discussed the different techniques used for feature extraction, representation, and classification of human activities. This paper\cite{bib18} reviewed the wearable sensors used for human activity recognition and the techniques used for feature extraction, representation, and classification. The authors also discussed the challenges and opportunities in this field. Deep learning models, such as recurrent neural networks (RNNs), convolutional neural networks (CNNs), and autoencoders, were used for this task. The authors found that the decision tree algorithm outperformed other algorithms in terms of accuracy. This study\cite{bib17} provides an overview of the current state of the art in deep learning for human activity recognition. The authors present a classification of deep learning models used for this task, including CNNs, RNNs, and Autoencoders, and discuss the challenges and opportunities in this field. The study also highlights the need for more robust evaluation metrics and benchmark datasets to advance the development of deep learning models for human activity recognition. Few studies show the improvement in optimisation methods like bio-inspired optimisations like Spider monkey optimisations \cite{kolkar2023human}. 

The research on ambient sensors and human activity recognition is carried out with PIR, temperature, sound and vision-based sensors. In\cite{bib19}, an accuracy of 98.4\% was achieved using a decision tree to combine multiple motion sensor features. However, adding power use, CO2, and sound sensors worsened the classification results. Here \cite{bib20}, the most recent developments of flexible and stretchable physical sensors are described, including temperature, pressure, strain sensors, and flexible and stretchable sensor-integrated platforms.

There are many datasets available publicly to analyse human activity recognition. WISDM (Wireless Sensor Data Mining) dataset\cite{bib21}: This dataset was created by researchers at Fordham University and contained data from over 3 million instances of human activities collected from wearable sensors. The activities include walking, jogging, sitting, standing, and stair climbing. The data is collected from smartphone accelerometers and gyroscopes and is freely available for research purposes. UCI HAR (Human Activity Recognition Using Smartphones) dataset\cite{bib22}: This dataset was created by researchers at the University of California, Irvine and contains data from experiments conducted with a group of 30 volunteers. The data includes 561 features extracted from smartphone accelerometers and gyroscope signals and is divided into test and train data. The activities in the dataset include walking, standing, sitting, lying, and various physical exercises. PAMAP2 (Physical Activity Monitoring) dataset\cite{bib23}: This dataset was created by researchers at the University of Karlsruhe and contained data collected from wearable sensors attached to the subjects' wrists, chests, and ankles. The data includes nine activities: walking, jogging, cycling, and jumping. The dataset is freely available for research purposes and is widely used for human activity recognition and pattern recognition tasks. HAPT (Human Activity and Posture Recognition) dataset\cite{bib24}: This dataset was created by researchers at the Technical University of Munich and contained data collected from wearable sensors attached to the subjects' waists. The data includes 26 activities: walking, running, jumping, and bending. The dataset is publicly available and is widely used for human activity recognition tasks. OPPORTUNITY (Opportunistic Sensing of Physical Activities) dataset\cite{bib25}: This dataset was created by researchers at the University of California, Berkeley and contained data collected from wearable sensors attached to the subjects' wrists, ankles, and pockets. The data includes 18 activities: walking, running, jumping, and sitting. The dataset is publicly available and is widely used for human activity recognition tasks. 

The datasets based on smartphones can recognise basic activities using deep learning techniques. The existing datasets based on smartphone-based sensors lack continuous data for days or weeks. In analysing human behaviour patterns based on smartphones, there is a need for a dataset with at least a month of data on a person. This motivated us to create a long-hour dataset. Also, hybrid sensor-based approaches in the literature helped us to combine smartphone-based sensors with ambient sensors like PIR and pressure sensors. The hybrid-based sensors allow for deriving more activities than the basic activities.

\section{Proposed Framework for Long-Hour Data Collection for Human Behavior Activity Pattern Recognition}

\begin{figure}[!ht]
	\centering
	\includegraphics[width=0.7\textwidth]{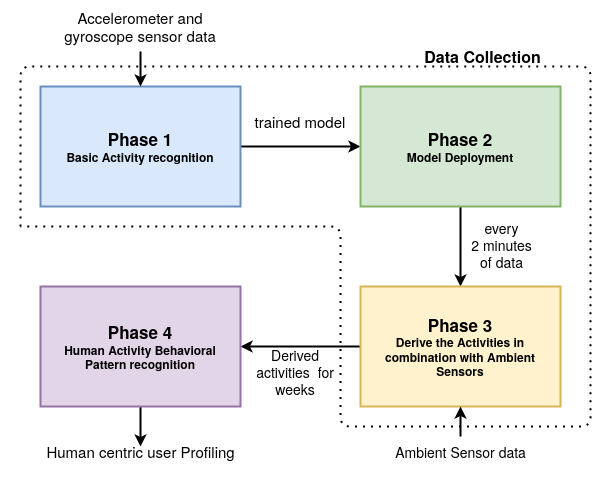}
	\caption{Proposed Framework for Long-Hour Data Collection for Human Behavior Activity Pattern Recognition.}
	\label{fig:phases_broad}
\end{figure}
\begin{figure}[!ht]
	\centering
	\includegraphics[width=0.7\textwidth]{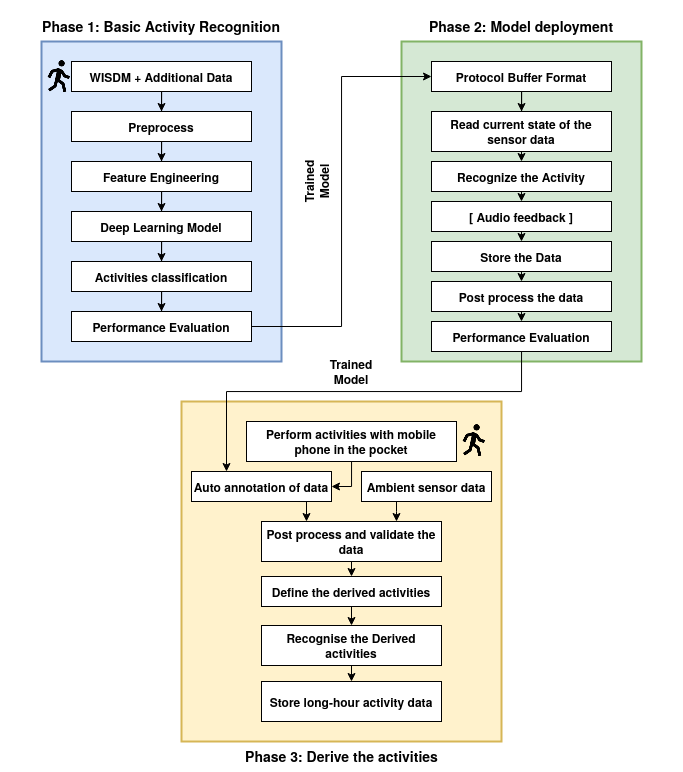}
	\caption{Phases of Long-Hour Data Collection for Human Behavior Activity Pattern Recognition.}
	\label{fig:phases}
\end{figure}

The long-hour dataset preparation is done in the first three phases, as shown in Fig \ref{fig:phases_broad}. In the first phase, the basic daily human activities like walking, sleeping, stair up, stairs down, and sitting are recognised using the deep learning model and are explained in detail \ref{subsec:phase1} and also shown in Fig \ref{fig:phases}. In Phase 2, the output of Phase 1 is taken as input to recognise the data. Here, the trained model from Phase 1 is deployed in the Android application to recognise and annotate the activities automatically. Phase 3 uses the Android application of Phase 2 to record the same activities over an extended time. Ambient sensors output data like location information and the appliance's ON and OFF status are also recorded. The derived activities are defined and recognised by combining the recognised basic activities and the ambient sensor-based data. At the end of all phases, the activities for the entire period of the day are recorded. The data recording is repeated for an average of 28 days from 5 subjects. 

The sensor placement is inside the trousers pocket, either the left or right pocket. The phone's orientation and the mobile inside the pocket status are considered and processed while creating and recognising the dataset. 
	
\subsection{Phase 1: Basic Human Activity Recognition}	
\label{subsec:phase1}

To recognize the basic activities, a WISDM data set is used. Wireless Sensor Data Mining lab (WISDM) Dataset\cite{bib21} is a publicly available laboratory-controlled human activity recognition dataset. The dataset uses 36 subjects with six activities: walking, sitting, jogging, downstairs, upstairs, and standing. The data is recorded using smartphones with inertial sensors like accelerometers and gyroscopes. The sampling rate of 20 Hz and the number of examples used are 1,098,207. There are a total of 6 attributes, and the class distribution is as follows; Walking: 424,400 (38.6\%), Jogging: 342,177 (31.2\%), Upstairs: 122,869 (11.2\%), Downstairs: 100,427 (9.1\%), Sitting: 59,939 (5.5\%), Standing: 48,395 (4.4\%). The data is later transformed into 10 seconds of accelerometer samples (200 records). The simple statistical measures are calculated to transform the data into 46 values(features) of a single row.

\begin{figure}[!hb]
	\centering
	\includegraphics[width=0.4\textwidth]{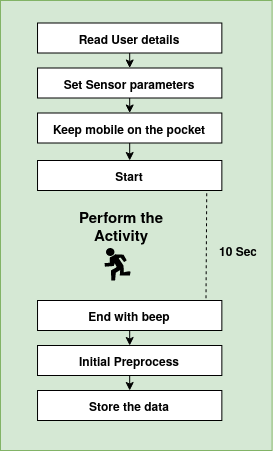}
	\caption{Process to collect the basic activities in Android application. }
	\label{fig:android_app_flow}
\end{figure}

However, it is observed that after deploying the model with the WISDM, the results could have been more accurate in the ground truth evaluation compared to the dataset. This might be because of the change in the smartphone sensor configuration and varied person-to-person walking styles. We experimented with data fusion at the data level with six more people data. These people are the subset with whom we are recording our long-hour dataset. The additional 64261 examples of data from 6 more people are added to the existing dataset. The same strategy is followed while recording the sensor data as in the original dataset to avoid data mismatch. We recorded the same activities as in the original and sampled every 50 ms(20 samples per second). The data is transformed into 10-second segments, and features are generated. Simple statistical methods like average, standard deviation, average absolute difference, average resultant acceleration, and time between peaks transform them into a single example/tuple of 46 values.

All the people were given Android mobile phones with the application installed. The instructions to record the activities are given prior. Users need to fill in the minimal data like name, age, and gender and agree to contribute to collecting the data. Users must select the activity they are performing, press the start button on the app, and complete the activity. The data collection team monitors the entire process of data collection. The user enters the start button and performs the activity selected. After 5 seconds beep indicates the recording has been started. The 5-second delay is given to keep the mobile phone inside the pocket and start performing the activity. The process to collect the basic activities in the Android application is shown in Fig \ref{fig:android_app_flow}. The team monitored and verified data collection with visualization and simple statistical methods. Each file is labelled with the subject ID  and the activity performed. Butterworth low-pass filters are applied to remove the noise, eliminating high-frequency components, if any, from the raw sensor data. 

\subsubsection{Preprocess}
The Smartphone sensors used in this study capture the changes in inertial movements. Accelerometer sensors measure acceleration forces, while gyroscope sensors measure angular velocity. By combining the data from these sensors, the system records the movement in the x, y, and z directions. To preprocess the input signals, nose filters are applied, and a sampling rate of 2.56 seconds is used. Butterworth low-pass filters are employed to remove noise from the data. The experiments are conducted using the WISDM datasets. The WISDM dataset consists of time series data that needs to be segmented into 128 timestamps per sample. Each data point in the dataset includes x, y, and z features, representing the sensor's direction. The accelerometer captures body and gravity acceleration, while the gyroscope records angular velocity and motion. Data cleaning involves removing duplicates and null values, and the dataset is balanced regarding class distribution.

\subsubsection{Feature Engineering}
For feature extraction, a combination of CNN and RNN architectures is utilized in this work to leverage the characteristics of time series data. The one-dimensional convolutional neural network (CNN) is particularly effective in extracting features directly from raw data, eliminating the need for manual feature extraction by domain experts. A feature map is generated as the output by applying filters to the input data. The activation function used in the CNN is the Rectified Linear Unit (ReLU), and kernel size filters are employed. Max pooling is applied to select the most important features from the previous feature maps, maximizing the elements in the feature map. This approach enables effective and automatic feature extraction without relying on prior domain knowledge.

\subsubsection{Deep Learning Models}

The Multi-layer Perceptron(MLP) computational model is used in this work that processes the raw data collected through interconnected computational nodes. These nodes are called neurons. These nodes are combined into many layers and combinations using connections with the weight. However, to apply time series data where the input and output are dependent on each other, Recurrent Neural Networks(RNN) are used. The activation of a neuron is fed back to itself with a weight and a unit time delay, which provides it with a memory (hidden value) of past activation, which allows it to learn the temporal dynamics of sequential data. These memory cells can be simple and complex with multiple parameter settings. RNN looks for all the previous nodes to compute the next state, which might lead to a vanishing gradient problem. There is a variation for RNN called long short-term memory(LSTM) networks, which selectively read, write and forget the information from previous steps.

\begin{figure}
	\centering
	\includegraphics[width=0.6\textwidth]{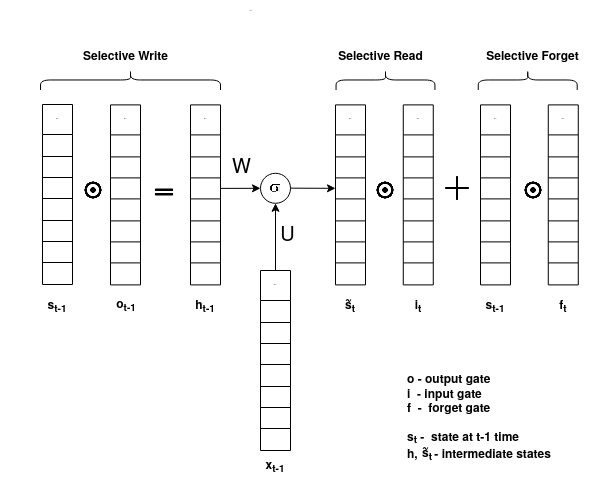}
	\caption{General working of the LSTM node.}
	\label{fig:LSTM_Explainations}
\end{figure}

In our time series dataset, the data in a window depends on the input of the previous data. LSTM layers help in extracting the temporal features in sequential data. The main work of the LSTM is by having three different gates, namely Input Gate- Equation \eqref{eqInputGate}, Output Gate- Equation \eqref{eqOutputgate}, and Forget Gate- Equation \eqref{eqForgetGate} also shown in the Fig \ref{fig:LSTM_Explainations}. The information $ x_t $ goes from state $ s_{t-1} $ to  $ s_t $ state- Equation \eqref{eqStateS}. Nevertheless, from the $ x_t $ information, we can selectively read, write and forget the information. In the selective write operation, only some data is written to the next state. The written portion should be significant to the model to reduce the loss. $ W $ and $ U $ matrix contains the weights of the input and recurrent connections. The sigmoid function is used to make the fraction between 0 and 1. 
\begin{equation}o_t = \sigma{(W_oh_{t-1} + U_ox_t + b_o)}  \label{eqOutputgate}\end{equation}
\begin{equation}i_t = \sigma{(W_ih_{t-1} + U_ix_t + b_i)}  \label{eqInputGate}\end{equation}
\begin{equation}f_t = \sigma{(W_fh_{t-1} + U_fx_t + b_f)}  \label{eqForgetGate}\end{equation}
\begin{equation}\Tilde{s}_t = \sigma{(Wh_{t-1} + Ux_t + b)}  \label{eqStateSTilda}\end{equation}
\begin{equation}s_t = f_t\odot s_{t-1}+i_t \odot \Tilde{s}_t  \label{eqStateS}\end{equation}
\begin{equation}h_t = o_t\odot \sigma({s_{t}})  \label{eqStateH}\end{equation}

Typical CNN architecture has the Convolution layer, poling layer, flattening and fully connected layers. This work uses CNN and RNN to use the time series data. One-dimensional convolution network performs well in feature extraction from raw data. That means there is no need for domain expertise to extract the features manually here. A feature map is obtained by obtaining the filters of the input data and collected as an output. In CNN, Relu(Rectified linear unit) is used as the activation function and kernel size filter. Max pooling is used to consider the essential features from the previous feature maps. This maximizes the elements in the feature map.

The LSTM network is the RNN composed of LSTM units. Gated Recurrent Units(GRUs) vary the LSTM networks and solve the vanishing gradient problem. The central aspect of GRU is it keeps the relevant information and passes it to the next steps. GRUs store and filter the information using the update and reset gates. The update gate decides what data to pass to the next step from the previous step. The update gate can be related to the input and forget gates of LSTM. The reset gate decides what data to forget to move to the next step. These two are essential parts of the GRU and solve the vanishing gradient problem of RNN. 

The Gated Recurrent Units(GRU) Model uses the gated mechanism in the Recurrent Neural Networks to recognize the activities. The GRU model is the variation of the LSTM model explained earlier. A detailed explanation of the data processing, the model implementation and its results and discussion are described in Section \ref{sec:Phase1_ImplementationAndResults}. It is observed that the accuracy for a model applied to WISDM data as a result of 82\%, and with the fusion data, the accuracy is 91\%.

\subsection{Phase 2: Model Deployment}

The Model Deployment section focuses on implementing and integrating the trained deep learning model into a practical application. This section details the steps taken to deploy the model, including converting the model into a format suitable for deployment, developing an Android application to capture sensor data, and post-processing the collected data. The goal is to create a robust and user-friendly system that can accurately recognize human activities in real time.

\begin{figure}[!hb]
	\centering
	\includegraphics[width=0.4\textwidth]{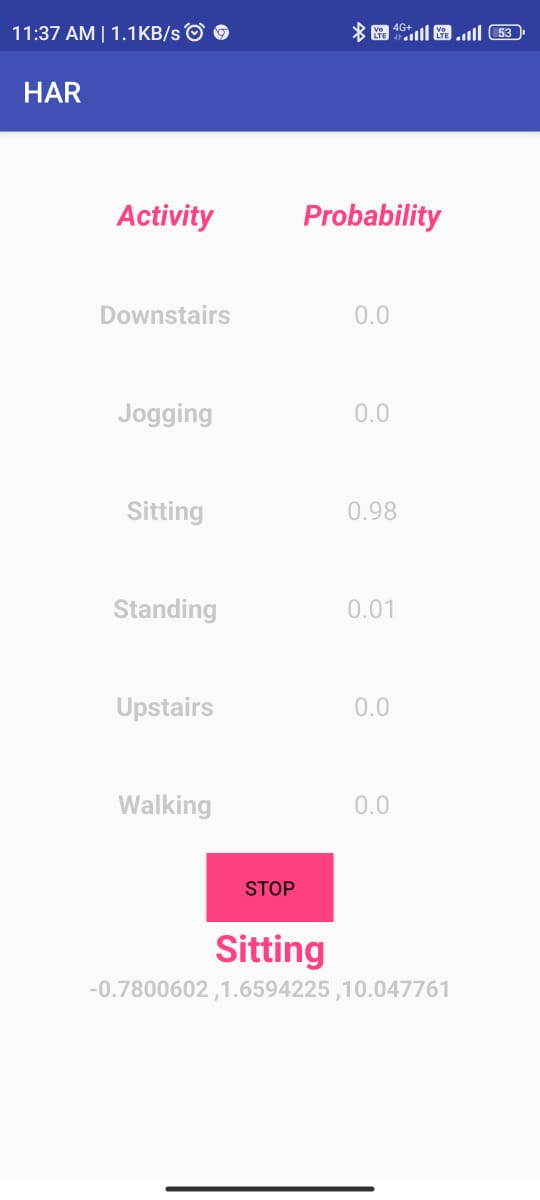}
	\includegraphics[width=0.4\textwidth]{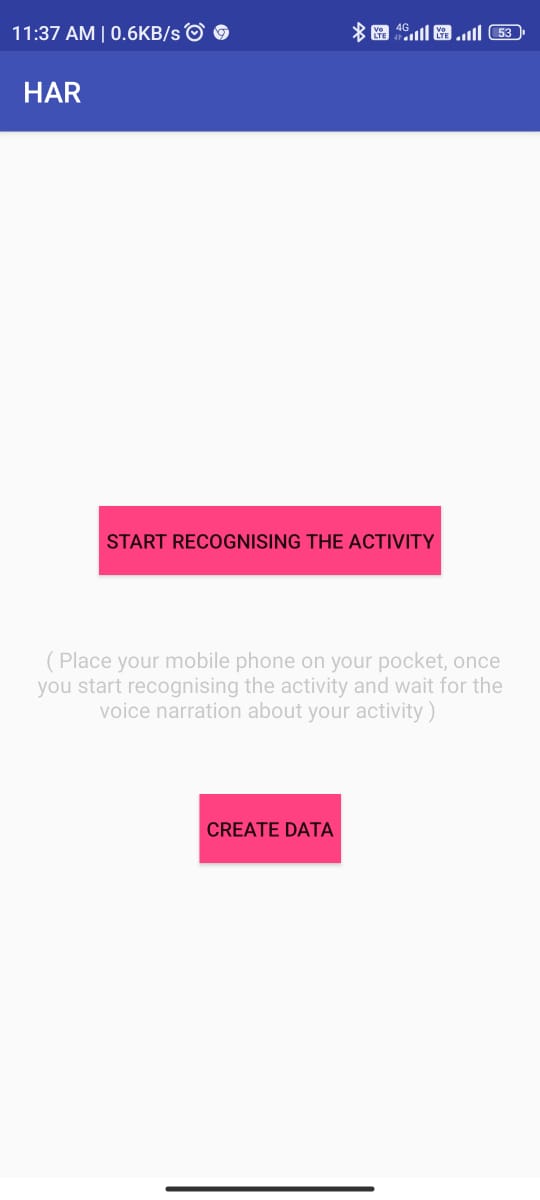}	
	\caption{ Sample screenshot of the Android application after deploying the trained model. The sample shows that walking activity is recognized.}
	\label{fig:app_screenshot}
\end{figure}

\subsubsection{Protocol Buffer Format}
We successfully deployed our deep learning model into our Android app by converting it into Protocol Buffers. This choice offers advantages such as optimized storage, streamlined communication, and enhanced compatibility. The compact binary representation of Protocol Buffers reduces the serialized model size, making it suitable for mobile devices with limited storage. The efficient communication between the app and the model ensures improved performance and responsiveness. Moreover, the language-agnostic nature of Protocol Buffers enables seamless integration with other server-side components or APIs. Overall, the utilization of Protocol Buffers has significantly contributed to efficiently integrating our deep learning model into the mobile environment.

\subsubsection{Read the Current State of the Sensor Data}

After reading the current states of the sensor data, including the accelerometer and gyroscope, we applied the same preprocessing strategy as during training. This involved filtering the raw sensor data, extracting relevant features, and normalizing the input. The preprocessed data was then fed into our trained deep learning model, which successfully recognized the activities it was trained for. By leveraging the power of deep learning algorithms and utilizing the processed sensor data, our model accurately classified and identified various human activities in real time. This demonstrated the effectiveness and applicability of our approach in activity recognition tasks.

\subsubsection{Recognise the Activities}

We have successfully recognized and classified six basic activities: sitting, standing, stair up, stair down, sleeping, Jogging and walking. Using our deep learning model and the analysis of sensor data from the accelerometer and gyroscope, we could accurately distinguish between these activities. Our model's training and validation processes enabled it to learn and generalize patterns associated with each activity, allowing for reliable recognition in real-world scenarios. 

\subsubsection{Audio Feedback and Data Storage}
The sensor data, comprising accelerometer and gyroscope readings, is stored in the mobile phone's storage. To facilitate seamless integration and analysis, we have implemented an automatic annotation process that assigns labels to the sensor data based on the recognized activities. By storing the sensor data and annotated labels, we ensure easy access for subsequent analysis, model training, and evaluation. Additionally, as an optional feature, we have incorporated audio feedback to verify the correctness of activity labels. This audio feedback serves as an additional means of validating the accuracy of the assigned labels and provides an interactive component to the recognition system. Users can listen to the feedback and verify the correctness of the recognized activities, enhancing the system's overall reliability.

\subsubsection{Post Processing the Data}
After training and building the model, we integrated it into our human activity recognition app, which offers several features. The "Sensor Settings" feature allows users to select the sensors to record data. The default sensor options and time intervals are set according to the objective of this work by the data collector, an expert in the domain. The "Profile Creation" module collects information about the subjects, such as their name, age, profession, and gender, providing contextual information for the experiment.

To optimize storage and efficiently manage the recorded data, we stored only one label per minute by taking the maximum of the activities recognized within that minute. Only the start and end times were noted for continuously repeating activities over a longer duration, significantly reducing the file size while preserving the relevant information. This file size reduction was crucial as we stored hours of data in the mobile phone memory, ensuring efficient use of available storage resources.

Figure \ref{fig:app_screenshot} presents a sample screenshot of the application, showcasing the recognition of the "Sitting" activity.
		
\subsection{Phase 3: Derive more activities by combining the ambient sensors}
In Phase 3 of our work, the main objective was to collect long-hour data to analyze further and derive new activities. Table \ref{table:Activities_List} presents the recorded activities, ambient sensor data, and the derived activities. The mobile activities, representing the basic activities, were auto-annotated and stored in text format, along with the corresponding ambient sensor data and timestamps. The synchronization between the two data types was straightforward since they were stored every minute. To ensure data quality, manual monitoring and validation were performed by the subjects and authors.

\begin{figure}[!ht]
	\centering
	\includegraphics[width=0.5\textwidth]{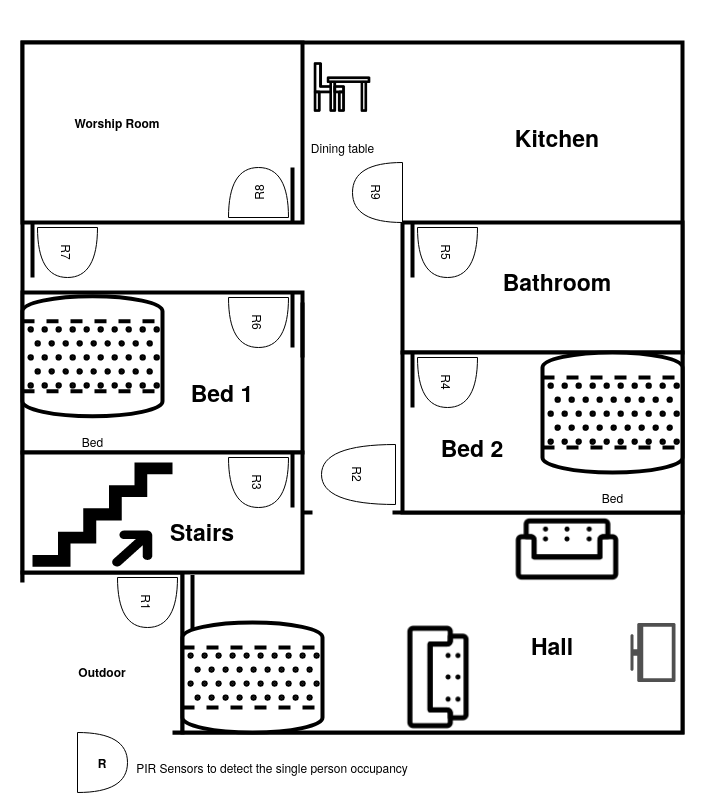}
	
	\caption{Floor design of the home in which the long-hour data is recorded. }
	\label{fig:floor_design}
\end{figure}
\begin{figure}[!ht]
	\centering
	\includegraphics[width=0.5\textwidth]{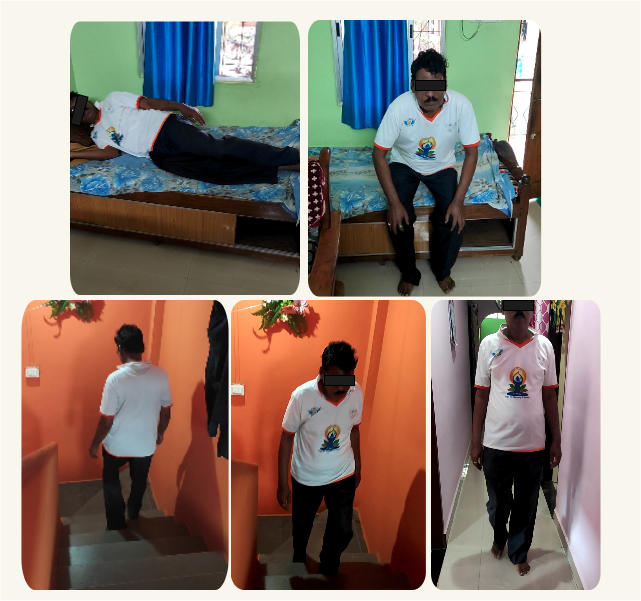}
	\caption{Samples of the basic activities performed by the subject. Shows Sleeping, Sitting, StairDown, StairUp, and Walking activities in order from Top Left to Bottom Right}
	\label{fig:sample_subject_image}
\end{figure}
\begin{figure}[!ht]
	\centering
	\includegraphics[width=0.5\textwidth]{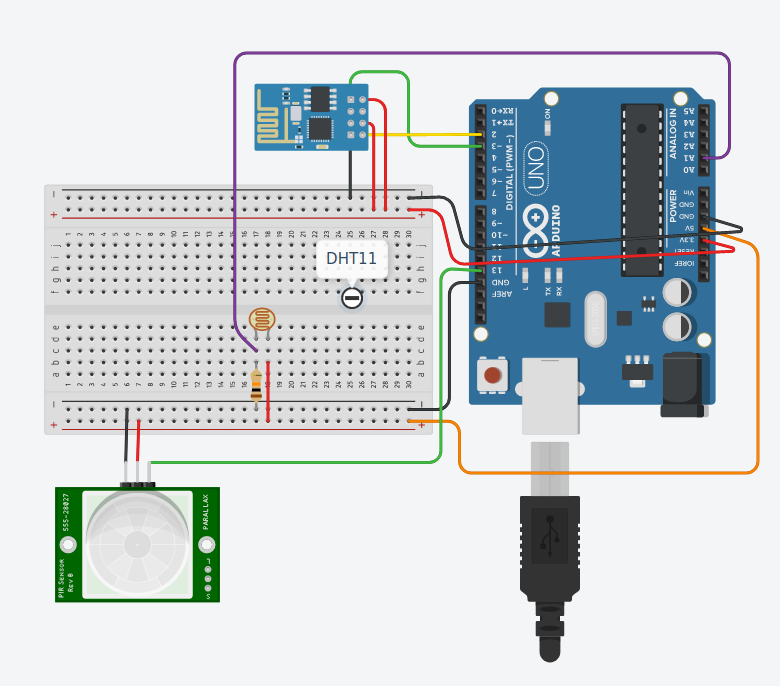}
	\caption{Circuit diagram for the single room occupancy detector using Arduino, PIR Sensor.}
	\label{fig:circuit}
\end{figure}

\subsubsection{Ambient Sensor Data}

To detect single occupancy in a room, we used PIR sensors placed at specific locations near the room's entrance, as depicted in Figure \ref{fig:floor_design}. The PIR sensor data was read and recorded using an Arduino microcontroller, with the data being stored on a micro SD card. The circuit diagram for this setup is shown in Figure \ref{fig:circuit}. The recorded data underwent post-processing to detect occupancy in the room.

The Algorithm \ref{alg:SPOD_MQTT} significantly determines the room's occupancy. Given that our work focuses on assisting elderly individuals living alone, we specifically considered single-person occupancy detection. The room context provided crucial information for deriving the data. The Single Person Occupancy Detector algorithm, identifies the person's room and determines the intervals during which the person was inside the room. A PIR sensor status of 1 indicates motion detection in the room, and the timestamp is recorded accordingly. Initially, the occupancy variable is set to 0, indicating that the person is not inside the room. When the PIR sensor status is 1, the occupancy is set to 1 until the status becomes 1 again. This algorithm enables the determination of intervals for each PIR sensor in all the rooms. Recording the actual clock time posed a challenge in our experimental setup, which we addressed by using the DS1307 RTC, a real-time clock sensor, to record the timestamps. However, a few assumptions were made during the sensor data recording process.
Furthermore, for the TV switch and mirror bulb, the ON/OFF status was recorded using a customized relay switch with the Arduino and the corresponding timestamps. As for the drinking activity, a simple force sensor was employed to record whether the bottle was picked up or not, indicating the action of drinking water. We create an instance of the OccupancyDetector class with the MQTT client ID, broker address, and topic as arguments. The paho.mqtt.client module is used to create an MQTT client object that subscribes to the specified topic and listens for updates from the sensor network. Inside the on\_message function, we get the current timestamp and read the sensor reading from the MQTT message payload. If the sensor reading is 1 and the occupancy status is 0, we update the occupancy status to 1 and append the current timestamp to the startTime list. If the sensor reading is 0 and the occupancy status is 1, we update the occupancy status to 0 and append the current timestamp to the endTime list.

\begin{algorithm}
	\caption{Single Person Occupancy Detector with MQTT}
	\label{alg:SPOD_MQTT}
	\begin{algorithmic}[1]
		\State Initialize client ID, broker address, and topic
		\State Initialize startTime and endTime as empty lists
		\State Create an MQTT client object
		\State Define the on\_connect function:
		
		\State Print the connection result code
		
		\State Define the on\_message function:
		
		\State Get the current timestamp
		\State Read the sensor reading from the MQTT message payload
		\If{sensor reading is 1 and occupancy status is 0}
		\State Update occupancy status to 1
		\State Append current timestamp to startTime list
		\ElsIf{sensor reading is 0 and occupancy status is 1}
		\State Update occupancy status to 0
		\State Append current timestamp to endTime list
		\EndIf
		
		\State Connect to the MQTT broker and subscribe to the topic
		\State Start the MQTT client loop
		\Function{get\_occupancy\_periods}{}
		\State \Return the startTime and endTime lists
		\EndFunction
	\end{algorithmic}
\end{algorithm}

\subsubsection{Post Processing and Data Validation}
In the post-processing stage of the data, we paid special attention to handling missing data. Any gaps or missing values in the sensor data were appropriately addressed to ensure the integrity of the dataset. Moreover, the label annotations assigned to the activities were carefully verified manually by the authors and cross-verified by the subjects themselves. This validation process helped ensure the accuracy and reliability of the activity labels associated with the sensor data. Two annotators manually corrected the errors, including the labelling and missing data, and their intra-rater reliability is Cohen's Kappa k= 0.90.

To facilitate further research on pattern recognition, we also focused on structuring the data in a suitable format. Proper data structuring allows for efficient data analysis and pattern extraction in our next work.

\subsubsection{Derived Activities}
There are two types of activities recorded: Basic Activities and Derived Activities. Basic Activities are Standing, Sitting, Walking, Jogging, Upstairs and Downstairs. Derived Activities are derived by combining the Basic Activities,  Ambient sensors like TV Switch ON/OFF, Mirror Bulb ON/OFF and considering the user's manual input. All the Activities are listed in Table \ref{table:Activities_List}.

\begin{table}[h]
	\begin{center}
		
		\caption{Derived activities and labelled activities using basic activity and ambient sensors for 2 minutes of data.}
		\label{table:Activities_List}%
		\begin{tabular}{@{}llll@{}}
			
			\toprule
			{Basic Activity} & {Ambient Sensor} & {Derived Activity} & {Labelled Activities}\\
			\midrule
			lie     & Bed Room         & Lying in Bedroom     & Lying in Bedroom     \\ 
			sit     & Bedroom          & Sitting in Bedroom   & Sitting in Bedroom   \\ 
			stand   & Bed Room         & Standing in Bedroom  & Standing in Bedroom  \\ 
			jog     & Bed Room         & Jogging in bedroom   & \textbf{Unnatural}            \\ 
			sleep   & Kitchen          & Sleeping in Kitchen  & \textbf{Anomaly}              \\ 
			lie     & Kitchen          & Lying in Kitchen     & \textbf{Unnatural}            \\ 
			sit     & Kitchen          & Sitting in Kitchen   & Sitting in Kitchen   \\ 
			stand   & Kitchen          & Standing in Kitchen  & Standing in Kitchen  \\ 
			jog     & Kitchen          & Jogging in kitchen   & \textbf{Unnatural}            \\ 
			lie     & Hall             & Lying in Hall        & Lying in Hall        \\ 
			sit     & Hall             & Sitting in Hall      & Sitting in Hall      \\ 
			stand   & Hall             & Standing in Hall     & Standing in Hall     \\ 
			jog     & Hall             & Jogging in Hall      & \textbf{Unnatural}            \\ 
			lie     & Worship          & Lying in Worship    & \textbf{Anomaly}              \\ 
			sit     & Worship          & Sitting in Worship   & Sitting in Worship   \\ 
			stand   & Worship          & Sanding in Worship   & Sanding in Worship   \\ 
			jog     & Worship          & Jogging in Worship   & Unnatural            \\ 
			lie     & Hall +TV         & Watching TV Lying    & Watching TV Lying    \\ 
			sit     & Hall +TV         & Watching TV Sitting  & Watching TV Sitting  \\ 
			stand   & Hall +TV         & Watching TV standing & Watching TV standing \\ 
			jog     & Hall +TV         & Watching TV Jogging  & Unnatural            \\ 
			stairUp & Stairs           & Stair Up             & Using Stairs         \\ 
			stairUp & Stairs           & Stair Down           & Using Stairs         \\ 
			-       & Outside          & Outside Activity     & Outside Activity     \\ 
			-       & WaterBottleUse   & Drinking Activity    & Drinking Activity    \\ 
			-       & MIrrorSwitchON   & Grooming             & Grooming             \\ 
			-       & BathroomSwitchON & BathroomActivity     & BathroomActivity     \\ 
			\botrule
		\end{tabular}
	\end{center}
\end{table}

The Derived activities use ambient sensors to label the activities. If the pressure sensor attached to the chair is active, the person is sitting on the chair. If the person is sitting on the chair and the TV is switched ON, then the person is watching the TV. Similarly, If the Bathroom switch is on, the person is using the bathroom. The table \ref{table:Activities_List} lists all the derived activities. Combining the recognized activity, location context and other ambient context derives more activities. An unnatural activity is an activity that is not possible in the regular context, and it is human-centric. For example, the Jogging in the kitchen is labelled as \textit{Unatural}. In some user contexts, this labelled activity might not be unnatural because people might sleep in the kitchen too. However, our dataset defines it as \textit{Unatural} because of the user's context. Anomaly is also an unnatural activity but in the context of the severity. 
	
	\section{Implementation and Result}
	This section briefs the experimental setup, implementation details and the results obtained. Each phase has a different methodology and is experimented with, one by one, in four phases.
	\label{sec:Phase1_ImplementationAndResults} 
	
	\subsection{Phase 1}
	
	This model has experimented on the WISDM dataset and the WISDM-fusion. The noise removal and the initial segmentation are carried out. The data is checked for missing data and filled using simple linear interpolation. The data is time-series and segmented to a sliding window of 128 with a 50\% overlap to preserve the temporal relationship.
	
	Feature extraction plays an essential role in classification. As it is time-series data, it has both temporal and spatial features. The preprocessed data is fed into the two layers of GRU with 32 neurons each to extract the temporal features, and CNN layers are used to extract the spatial features. The CNN layers have experimented with multiple layers, and in this work, 32 and 128  filters with a kernel size of 64 each is used in the CNN layers. The dropouts of 0.07 and 0.2 are used after CNN and GRU layers. The details of other hyperparameters are listed in Table \ref{table:hyper parameters}. The following are the steps in building the GRU model for classification activities. 
	
	\begin{enumerate}
		
		\item Input: read the raw data, read from sensors. The vector size is 128.		
		\item CNN layers: Convolution operations of 1 dimension are applied on input. The Relu Activation function is used with this operation. Another convolution operation is followed by the first.
		\item Dropout: This layer is added to avoid over-fitting problems in a multi-layer neural network. The dropout of 0.07 is used.
		\item Max pooling: As there are multiple layers, the pooling layer separately operates upon each feature map to create a new set of the same pooled feature maps to order the CNN layers.
		\item Flatten: This step is necessary to make a proper input vector to the next GRU layer.
		\item GRU layers: With 32 units, it is again fed to another GRU layer
		\item Dense: The output of all the above layers is concatenated using a dense layer.
		\item Softmax Activation: This is the final layer for outputting the multiple class activities. This layer follows the batch normalization of all the layers.
	\end{enumerate}
	
	\begin{table}[h]
		\begin{center}
			\caption{Hyper parameters settings for GRU model.}
			\label{table:hyper parameters}
			
			\begin{tabular}{@{}ll@{}}
				\toprule
				\textbf{Hyper Parameters} & \textbf{Value}           \\ 
				\midrule 
				Conv filters 1            & 32                \\ 
				Conv filters 2            & 128    \\ 
				Dropout 1                 & 0.07         \\ 
				GRU Units 1               & 64         \\ 
				GRU Units 2               & 64                        \\
				Dropout 2                 & 0.2                       \\
				Epochs                    & 50                        \\
				Batch size                & 64                        \\
				Loss                      & categorical\_loss entropy \\
				Optimiser                 & adam                      \\
				Learning rate             & 0.001                     \\
				\botrule
			\end{tabular}
		\end{center}
	\end{table}

	\subsubsection{Result and Discussion}
	
\begin{figure}[!ht]
		\centering
		\includegraphics[width=0.8\textwidth]{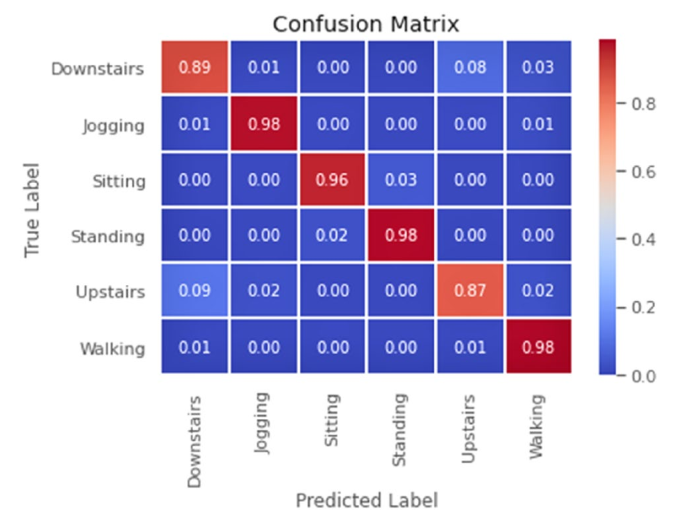}
		\caption{Confusion matrix of GRU model experimented on WISDM Dataset.}
		\label{fig:confusionMatrix}
	\end{figure}

\begin{figure}[!ht]
		\centering
		\includegraphics[width=1\textwidth]{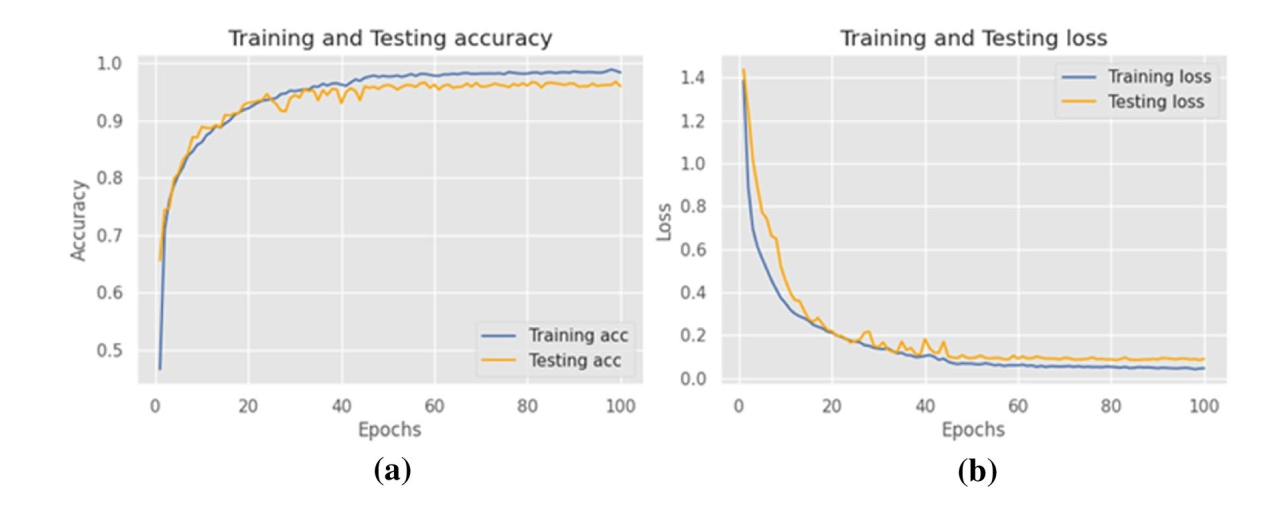}
		\caption{Train and testing accuracies and Training and Testing Loss for WISDM Dataset.}
		\label{fig:train_test_Loss}
	\end{figure}

	In the first phase of work, we evaluated the WISDM dataset and calculated the model's accuracy before deployment, shown in Table \ref{table:accuracy_BeforeDeployment}. There are a total of 36 user data present in the original dataset. Initially, we took out eight users from the dataset for test data and trained the model with the remaining data. Figure \ref{fig:confusionMatrix} shows the confusion matrix with the highest accuracy for standing and walking activity. This is because there is fewer true negative for this data. Figure \ref{fig:train_test_Loss} shows the Train and Testing accuracy and the Training and Testing Loss. 
	
	It is observed from Table \ref{table:accuracy_AfterDeployment} that the accuracy is lesser after deploying the trained model on the mobile phone. This is because the environment in which the WISDM data is created differs from the deployment environment. Even the users involved in testing after the deployment is varied because of the varied human style of performing certain activities. So, we added 11 users' data to the existing dataset to make WISDM Fusion data. The data level fusion is done. The experiment setup follows a similar preprocessing strategy used in the original dataset. The dataset is trained again and tested with the same model. It is observed from Table \ref{table:accuracy_AfterDeployment} that the accuracy is increased after adding new users' data(WISDM Fusion). We did add the new users not only to improve the accuracy of the data in the new environment but also to use the same users while creating the long-hour dataset to collect the data for human pattern recognition. 
	
	\begin{sidewaystable}
		\sidewaystablefn%
		
		\begin{center}
			\caption{Accuracy of GRU model experimented on WISDM and WISDM Fusion data before deployment.}
			\label{table:accuracy_BeforeDeployment}
			\begin{tabular}{@{}lcc@{}}
				\toprule
				\multirow{2}{*}{\textbf{Activities}}     & \multicolumn{2}{c}{\textbf{Accuracy}}\\ \cline{2-3}                       
				& \multicolumn{1}{l}{\textbf{WISDM}} & \textbf{WISDM Fusion} \\
				\midrule 
				
				Walking                              & \multicolumn{1}{c}{0.98}           & 0.99                  \\
				Sitting                              & \multicolumn{1}{c}{0.98}           & 0.97                  \\
				Upstairs                             & \multicolumn{1}{c}{0.87}           & 0.89                  \\ 
				Downstairs                           & \multicolumn{1}{c}{0.89}           & 0.91                  \\ 
				Jogging                              & \multicolumn{1}{c}{0.96}           & 0.97                  \\ 
				Standing                             & \multicolumn{1}{c}{0.98}           & 0.98                  \\ 
				\textbf{Average}                     & \multicolumn{1}{c}{\textbf{0.94}}  & \textbf{0.95}         \\ 
				
				\botrule
			\end{tabular}
		\end{center}
		\vspace{10mm}
		\begin{center}
			\begin{minipage}{\textheight}
				\caption{Accuracy of GRU model experimented on WISDM and WISDM Fusion data after deployment.}
				\label{table:accuracy_AfterDeployment}
				\begin{tabular*}{\textheight}{@{\extracolsep{\fill}}lcccccc@{\extracolsep{\fill}}}
					\toprule
					\textbf{Activities} & \begin{tabular}[x]{@{}c@{}}\textbf{Activity data}\\ \textbf{in seconds}\end{tabular}   & \begin{tabular}[x]{@{}c@{}}\textbf{Total} \\ \textbf{Activities}\end{tabular}  & \begin{tabular}[x]{@{}c@{}} \textbf{Correctly}  \\ \textbf{Recognised} \end{tabular}& \begin{tabular}[x]{@{}c@{}} \textbf{Accuracy} \\ \textbf{trained model} \\\textbf{on WISDM}\end{tabular} &\begin{tabular}[x]{@{}c@{}} \textbf{Correctly}  \\ \textbf{Recognised} \end{tabular} & \begin{tabular}[x]{@{}c@{}}\textbf{Accuracy} \\\textbf{trained model} \\\textbf{on WISDM Fusion} \end{tabular}\\ 
					\midrule 
					Walking & 2356 & 471 & 461     & 0.98  & 453     & 0.96         \\ 
					Sitting             & 1866      & 373 & 352     & 0.94  & 362     & 0.97         \\ 
					Upstairs            & 1362      & 272 & 232     & 0.85  & 235     & 0.86         \\ 
					Downstairs          & 1536      & 307 & 270     & 0.88  & 275     & 0.90         \\ 
					Jogging             & 2653      & 530 & 489     & 0.92  & 506     & 0.95         \\ 
					Standing            & 962       & 192 & 185     & 0.96  & 183     & 0.95         \\ 
					\textbf{Total/Average}    &
					
					10735.00		
					& 2145.00           & 1989.00   & \textbf{0.92}       & 2014.00
					& \textbf{0.93}   \\ 
				\end{tabular*}
			\end{minipage}
		\end{center}
		
	\end{sidewaystable}

	\begin{sidewaystable}
		\sidewaystablefn%
		\vspace{10mm}
		\begin{center}
			\caption{Validation after Deployment. Confusion Matrix shows the actual values of the recording and App output values for the WISDM Dataset.}
			\label{table:Manual_confusion_Wisdm}
			\begin{tabular}{@{}ccccccccc@{}}
				\toprule
				
				& \textbf{walk} & \textbf{sit} & \textbf{stairup} & \textbf{stair down} & \textbf{jog/run} & \textbf{sleep} & \textbf{stand} & \textbf{recongnised}\\ 
				
				\midrule 
				
				\textbf{walk}       & 426 & 2   & 9   & 11  & 23  & 0   & 0   & 471 \\ 
				\textbf{sit}        & 6   & 336 & 2   & 2   & 1   & 11  & 15  & 373 \\ 
				\textbf{stairup}    & 11  & 2   & 195 & 18  & 42  & 1   & 3   & 272 \\ 
				\textbf{stair down} & 18  & 2   & 25  & 239 & 14  & 0   & 9   & 307 \\ 
				\textbf{jog/run}    & 8   & 2   & 15  & 34  & 456 & 0   & 15  & 530 \\ 
				\textbf{sleep}      & 1   & 13  & 0   & 0   & 0   & 457 & 1   & 472 \\ 
				\textbf{stand}      & 3   & 6   & 0   & 0   & 0   & 5   & 178 & 192 \\ 
				\botrule
			\end{tabular}
		\end{center}
		\begin{center}
			\caption{Validation after Deployment. Confusion Matrix shows the actual values of the recording and App output values for the WISDM Dataset.}
			\label{table:Manual_confusion_Wisdm}
			\begin{tabular}{@{}ccccccccc@{}}
				\toprule
				
				& \textbf{walk} & \textbf{sit} & \textbf{stairup} & \textbf{stair down} & \textbf{jog/run} & \textbf{sleep} & \textbf{stand} & \textbf{recongnised}\\ 
				
				\midrule 
				
				walk       & 453  &   0  & 2       & 2          & 14      &     0  &   0    & 471         \\ 
				sit        & 2    & 362 &      0   &       0     &    0     & 2     & 7     & 373         \\ 
				stairup    & 5    &  0   & 213     & 16         & 38      &   0    &    0   & 272         \\ 
				stair down & 12   &  0   & 42      & 245        & 8       &     0  &   0    & 307         \\ 
				jog/run    & 5    &   0  & 8       & 11         & 506     &    0   &    0   & 530         \\ 
				sleep      &  0    & 9   &  0       &      0      &   0      & 462   & 1     & 472         \\ 
				stand      & 3    & 6   &   0      &    0        &   0      &    0   & 183   & 192         \\ 
				\botrule
			\end{tabular}
		\end{center}
		
	\end{sidewaystable}

	\subsection{Phase 2} 
	
	The trained model from Phase 1 is deployed in the Android-based mobile phone. The protocol buffer format from Python is used to deploy the .pb file in the mobile application. Even though the trained model gave a better result of 94\% with the WISDM, It was challenging to achieve the same result after the deployment. This is mainly because the testing environment and user differed from the original dataset. 
	We analyzed the performance of the deployed version of the model from the smartphone. Each user performed the activities while individually keeping the mobile phone in the pocket. The automatic annotation from the trained model is stored in a file against each activity. The number of currently recognized activities from each file is saved in a file. Also, audio feedback is given to the user about the recognized activity. The audio feedback helps immediately with the correctly recognized activity. This is repeated for all users. From Table \ref{table:accuracy_AfterDeployment}, it is observed that a total of 10735 seconds of data is tested, and out of the 2145 activities, 1989 activities are correctly recognized using the trained model from WISDM data. The accuracy of 93\% is achieved after the training model's deployment on WISDM Fusion data.

	\subsection{Phase 3}

	In phase 3, the long-hour data is recorded. The activities recognized from the mobile app are stored in the file and the ambient sensor data. We took every 2 minutes of data and labelled the activity by following the algorithm \ref{alg:2_min_label}. The significance of accomplishing the two-minute profiling is reducing the number of activities for a long-hour dataset for human activity pattern recognition and not compromising the crucial activities considering the entire day of activity data. The window size of the profiling can be varied to Five-Minute profiling or ten-Minute profiling, depending on the application that is developed using human patterns. Table \ref{table:Example_10_min_Profiling} shows the sample instance of ten-minute labelling. The labels are recorded every 5 seconds in the phone's memory. There is a total of 24 entries of labels in the 2 minutes of data. This 2 minutes of data is passed to the algorithm to get one label every 2 minutes. The priority list will consist of the list of activities in priority. The significance of the priority list is to label the activity of importance, not just by the frequency. For example, consider a scenario where walking activity has the highest frequency in a two-minute window. However, the drinking activity has a lesser frequency in that two-minute window. We cannot neglect drinking activity just because the frequency is less. So, the drinking activity's priority will be more than the walking activity in the two-minute profiling. The Algorithm \ref{alg:2_min_label} checks the activities from the priority list and the frequency of all the activities. The priority of the activity is higher than that activity is labelled, or the activity with the higher frequency is labelled. The simple case of a priority setting for the activities is shown in Table \ref{table:sample_priority}. 
	
	The priority is set while labelling the activity for every interval. For example, in 5-minute profiling, taking the maximum occurred activity as a label is not a good criterion. This is because some activities rarely happen in 5 minutes. For example, the Drinking water activity is just for a maximum of 30 seconds, but this is a crucial activity to e considered while creating the Five-minute labelling for a day. So, the priority needs to be set for the drinking activity while labelling the 5-minute, and the rarely-occurring fall activity must be addressed. The priority is set according to the following equation.
	\begin{center}	
		let $ s = \{A_1,A_2,A_3,.. A_k\}$ where $A_i$ are the activity list\\		
		Priority of $A_i>=A_j $ where $Freq(A_i)<=Freq(A_j)$	
	\end{center}
	The priorities for the lesser frequency equations are set at a higher frequency. However, the user-centric priority list can be manually set depending on the requirement. 
	
	Now, more activities are derived by combining the two-minute labelled data and the corresponding ambient sensors data. Table \ref{table:Activities_List} lists the derived activities. Some activities are labelled as unnatural. For example, Lying in the Kitchen. Lying in the kitchen activity is not possible in a usual scenario. However, all the unnatural labels defined in our work are user-specific. Sleeping is an activity which is derived from lying activity. The duration of lying is more than 5 minutes without any movement is considered sleeping. Sleeping in the Kitchen activity is derived as an anomaly. We define an anomaly as unnatural activities, but the effect of that activity is severe and needs to be attained. For example, Sleeping in the Kitchen might be a fall and unconscious activity.
	
	\begin{algorithm}
		\caption{Two-Minute Labelling Strategy}
		\label{alg:2_min_label}
		\begin{algorithmic}[1]
			
			\Require data(timestamp,labels), PriorityList
			\Ensure  label			
			
			\State FrequencyList[]= getFrequency(labels);
			\If { $set(PriorityList).isdisjoint(FrequencyList)$}
			\If {$count(frequency(label))>1$}
			\State{$label = last(FrequencyList)$}
			\Else
			\State {$label =  first.FrequencyList$}
			\EndIf
			\Else
			\State {$label =  first.PriorityList$}			
			\EndIf
			
			\State 	\Return $label$ 
			
		\end{algorithmic}
	\end{algorithm}

	\begin{table}[]
		\caption{Simple case of priority setting for the activities. The smallest number indicates the highest priority.}
		\label{table:sample_priority}
		\centering
		
		\begin{tabular}{cc}
			\hline
			\multicolumn{1}{c}{\textbf{Activity}} & \textbf{Priority} \\ \hline
			Drinking Activity & 1        \\ 
			Sitting in Hall   & 2        \\ 
			Kitchen Activity  & 2        \\ 
			Walking Outside   & 3        \\ \hline
		\end{tabular}%
		
	\end{table}

	
	\begin{landscape}
		
		\begin{table}[]
			\caption{Sample instance of labelling for a 10-minute window using priority.}
			\label{table:Example_10_min_Profiling}
			
			\begin{center}
				\begin{tabular}{@{}ccc@{}}
					\hline
					\textbf{cases, operation} &
					\textbf{Input} &
					\textbf{10-minute labelling} \\ \hline
					\multirow{5}{*}{\begin{tabular}[c]{@{}c@{}}no priority,\\  Max()\end{tabular}} &
					1 Walking outside &
					\multirow{5}{*}{Walking Outside} \\ 
					& 2 Walking outside                       &  \\ 
					& 3 Walking outside                       &  \\ 
					& 4 Walking outside                       &  \\
					& 5 Walking outside                       &  \\ \hline
					\multirow{5}{*}{\begin{tabular}[c]{@{}c@{}}priority, \\ \\ HighestPriority()\end{tabular}} &
					1 Sitting in Hall &
					\multirow{5}{*}{Drinking Water} \\ 
					& 2 Sitting in Hall                       &  \\ 
					& 3 Sitting in Hall                       &  \\ 
					& 4 Kitchen Activity                      &  \\ 
					& 5 Drinking water                        &  \\ \hline
					\multirow{5}{*}{\begin{tabular}[c]{@{}c@{}}equal frequency,\\ equal priority, \\ \\ ChoseOccurredFirst()\end{tabular}} &
					1 Sitting in Hall &
					\multirow{5}{*}{Kitchen Activity} \\ 
					& 2 Sitting in Hall  &  \\
					& 3 Kitchen Activity &  \\
					& 4 Kitchen Activity &  \\
					& 5 walking outside  &  \\ \hline
				\end{tabular}%
			\end{center}
		\end{table}
		
		\begin{table}[]
			\caption{Comparison of Existing Datasets with the Proposed Dataset}
			\label{table:comparison_datasets}
			\resizebox{\columnwidth}{!}{%
				\begin{tabular}{cccccccccr}
					\hline
					\textbf{Dataset} & \textbf{Duration} & \textbf{\begin{tabular}[c]{@{}c@{}}Activities\\  Detected\end{tabular}} & \textbf{\begin{tabular}[c]{@{}c@{}}Behaviour \\ or \\ Habit \\ Identifiaction\end{tabular}} & \textbf{Subjects} & \textbf{Sensors} & \textbf{\begin{tabular}[c]{@{}c@{}}Ambient \\ Sensors\end{tabular}} & \textbf{\begin{tabular}[c]{@{}c@{}}Activity Recognition \\ Model\end{tabular}} & \textbf{Accuracy (\%)} & \multicolumn{1}{l}{\textbf{Year}} \\ \hline

					\begin{tabular}[c]{@{}c@{}}\textbf{Proposed} \\ \textbf{Dataset}\end{tabular} & 3 months & Daily activities (21) & yes & 12 & \begin{tabular}[c]{@{}c@{}}Smartphone (Accelerometer, \\ Gyroscope), Ambient Sensors\end{tabular} & Yes & LSTM, GRU & 95 & \multicolumn{1}{l}{N/A} \\
					
					CASAS (\cite{casas}) & 15 days & Daily activities & yes & 10 & \begin{tabular}[c]{@{}c@{}}Multimodal (Accelerometer, \\ PIR sensors, etc.)\end{tabular} & Yes & HMMs, CRFs & 91.2 & 2018 \\
					
					ExtraSensory \cite{extra_sensory} & (2 - 25 days) & \begin{tabular}[c]{@{}c@{}}Various activities\\ (116)\end{tabular} & yes & 60 & \begin{tabular}[c]{@{}c@{}}Smartphone (Accelerometer, \\ Gyroscope), Wearable sensors\end{tabular} & Yes & DNN & 83.6 & 2015 \\
					
					SisFall \cite{sisfall} & Within a day & 19 activities & No & 22 & Smartphone accelerometer & No & Naive Bayes & 95.8 & 2014 \\
					
					UCI-HAR\cite{bib22} & 6 hours & 6 activities & No & 30 & Accelerometer, Gyroscope & No & Deep Neural Network & 95.6 & 2012 \\
					
					WISDM\cite{bib21} & 2 weeks & 5 activities & No & 29 & Accelerometer & No & Decision Tree & 89.4 & 2010 \\
					
					OPPORTUNITY\cite{bib25} & 48 hours & 19 activities & No & 4 & Accelerometer, Gyroscope, etc. & No & Conditional Random Fields & 90.1 F1-score & 2012 \\
					
					PAMAP2\cite{bib23} & 5 hours & 12 activities & No & 9 & Accelerometer, Gyroscope, etc. & No & Random Forest & 82.4 & 2012 \\
					
					MobiFall\cite{mobifall} & Variable & 17 activities & No & 26 & Accelerometer & No & Support Vector Machine & 85.6 F1-score & 2013
				\end{tabular}%
			}
		\end{table}
	\end{landscape}
	
\subsection{Comparison with existing datasets}

Our proposed dataset is a comprehensive and valuable resource for activity recognition and behaviour/habit identification in the context of elderly monitoring, as mentioned in Table \ref{table:comparison_datasets}. For three months, it captures an extensive range of daily activities (21) and involves a diverse group of 12 subjects. What sets our dataset apart is the utilization of multimodal sensors, including smartphone sensors such as accelerometers and gyroscopes, along with ambient sensors like PIR, smart switch, and weight sensors. Combining these sensors gives us a holistic view of the participants' activities and environmental context.

Regarding activity recognition models, our approach leverages LSTM and GRU models, which have demonstrated excellent performance in sequence-based data analysis. Our accuracy of 87.3\% showcases the effectiveness of our proposed approach in accurately recognizing and classifying daily activities. Moreover, our dataset includes behaviour/habit identification, allowing for deeper insights into long-term patterns and routines.

Comparing our work to existing datasets, such as CASAS and ExtraSensory, our dataset offers a longer duration and a more diverse range of activities. Including ambient sensors further enriches the contextual information, providing a better understanding of the participants' living environment. Additionally, our accuracy results are competitive with those of other datasets, highlighting the reliability and effectiveness of our activity recognition models.

Overall, our proposed dataset not only addresses the specific needs of elderly monitoring but also presents a comprehensive and well-curated resource for activity recognition and behaviour/habit identification research. Its unique combination of multimodal sensors, extended duration, and diverse activities sets it apart from existing datasets and opens new avenues for advanced analysis and understanding of elderly behaviours.

\section {Human Behavioural Pattern Recongition}

In phase 4, the derived activities are noted for a week by a specific user, and user profiling can be created. User profiling helps in understanding the patterns of activities for a month or more and take a constructive decision based on the outcomes of the profile. For example, if the walking duration over the week is reduced, the user can take a proper decision on his walking duration for next week onwards. 
The pattern recognition for the user is done in varied intervals of profiling. Two-minute daily labelling provides more granular activities than the Five-minute labelling strategy. Two, five, and ten-minute labelling data are used and analyzed while creating the patterns. Also, the activity patterns can be derived daily or weekly.  

\begin{figure}[t]
	\centering
	\includegraphics[width=1\textwidth]{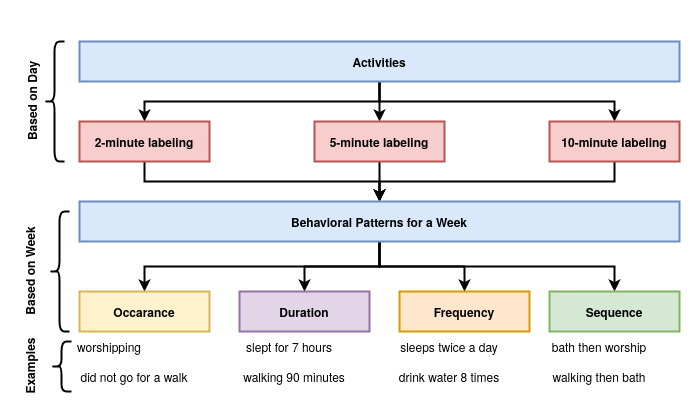}
	\caption{Methadology for Pattern Recognition.}
	\label{fig:pattern_methodology}
\end{figure}

\subsection{Day Profiling}
In the day profiling, the day-wise user activities are recognized. However, defining the 2-minute labelling will not give insight into human behaviour. The profiling can be derived from activity-oriented or interval-oriented. In activity-oriented profiling, the profiling is done on specific activities. For example, sleeping activity-specific patterns can be derived. For example, a person's sleep cycle can be derived from the number of hours the user sleeps in a week or the number of times the user sleeps in a day. The specified interval can be considered for deriving the patterns in interval-oriented activities. For example, Drinking activity is only four times in the morning(06:00 AM to 11:00 AM). Worship activity is not done in the evening. 

\subsection{Weekly Routines}
In Weekly routing, the data is calculated for every week. The weekly routine is done based on the following four methods.
\begin{enumerate}
	\item Occarance-based: Each activity recognized in the 2-minute, 5-minute, and 10-minute are considered in the occurrence-based profiling. It will check whether the activity has occurred or not in a day. For example, an activity might be done on one day, not another. For example, walking is done on D1 but not on D2. 
	\item Duration-based: Each activity recognized in the 2-minute, 5-minute, and 10-minute are considered in the duration-based profiling. It will check whether the activity has occurred for how much duration in a day. For example, sleeping is done for 6 hours on D1 and 8 hours on D2. The instance of the activity duration-based pattern is shown in Figure \ref{fig:PieChart_day} and \ref{fig:Chart_weekely}.
	
	\begin{figure}[!ht]
		\centering
		\includegraphics[width=1\textwidth]{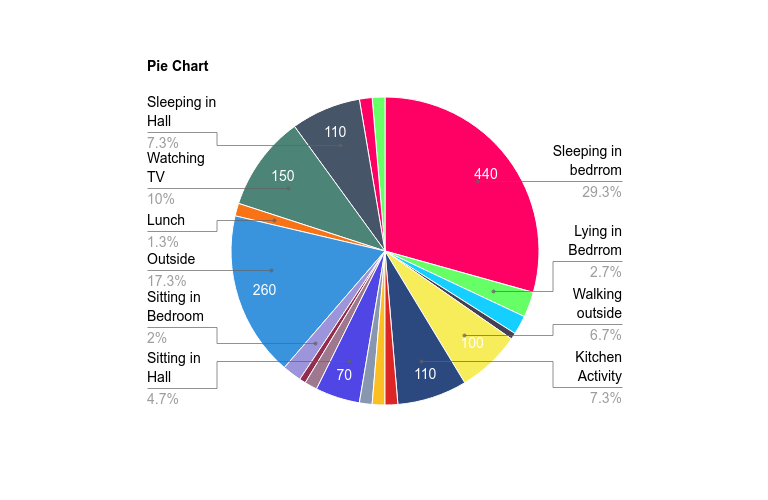}
		\caption{Duration-based analysis of activities for an entire day.}
		\label{fig:PieChart_day}
	\end{figure}
	\begin{figure}[!ht]
		\centering
		\includegraphics[width=1\textwidth]{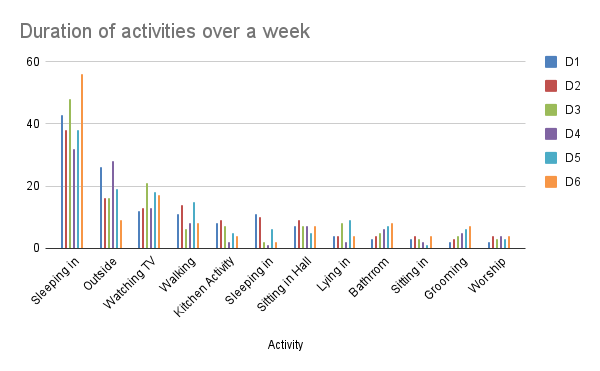}
		\caption{Duration-based analysis of various activities for a week.}
		\label{fig:Chart_weekely}
	\end{figure}
	\item Frequency-based: This will give insight into how often the activity is repeated daily. For Example, The sleeping activity is done thrice in D1 and twice in D2. 
	\item Sequence-based: This is the complex pattern recognition based on the sequence of the activities. The intervals have been defined to list the sequence of the pattern. The interval can be half-hourly, hourly, or quarterly in a day. For every interval, there will be a sequence of activities that happened. We look for the same pattern sequence over a week in this problem. 
	
	Only duration-based activity patterns are listed as a baseline human activity pattern recognition. The study and experiment of frequency-based and sequence-based activity pattern analysis are not carried out in this work. This is ongoing work, and we have restricted our work only to duration-based pattern recognition for this article. 
	
\end{enumerate}

\section{Constraints and Limitations}

This section discusses the constraints and limitations associated with our research.

\subsection{Sensors and the Activities}

Our research uses smartphone-based sensors such as accelerometers and gyroscopes to detect human activities. While this limits the range of activities we can recognize, our broader goal is to develop a comprehensive framework for human behaviour pattern recognition using this data. We have focused on a minimum set of activities as a starting point, but there is potential to expand and improve upon them in future iterations.

It is reasonable to state that variations in smartphone sensor configurations can impact the system's accuracy. However, the potential effects of sensor variations are minimized by ensuring the use of the same sampling rates and filters across devices. Even in our research, we experimented with multiple devices from different companies, and the accuracies remained consistent. We also implemented post-processing scripts to address missing values, nan, and calibration errors and performed manual verification and validation to ensure data integrity.

Our target audience is elderly people within the context of smart homes and the activities of daily living. While our system may not currently cater to activities such as cycling, specific kitchen activities, or mountain climbing, which may require additional sensors or specialized algorithms, our proposed system aims to provide reliable recognition for the selected basic activities. Including basic activities allows for comprehensive research on long-hour pattern detection and user profile creation based on these activities. We intend further to expand the activity repertoire in our ongoing research.

Individual variations in walking styles can impact the system's accuracy; our focus is primarily on distinguishing basic activities such as walking, sitting, standing, lying down, and stair-related movements. The system aims to identify the overall walking pattern rather than specifically differentiating between various walking styles. Regardless of individual variations, walking can still be recognized as a distinct activity compared to other activities. However, it is essential to acknowledge that extreme cases, such as individuals whose walking style closely resembles running, can present challenges that must be addressed for accurate activity recognition.

This work assumes the presence of a single person in the room during activity detection. This motivation stems from our aim to develop applications for the remote monitoring of elderly individuals who live alone. By focusing on single-person occupancy, we can tailor our system to address the specific needs and challenges associated with ensuring their safety and well-being.

\subsection{Privacy}

Regarding privacy, it is important to note that our system does not utilize GPS locations. Instead, we rely on identifying the location of the room in which the individual resides. This approach helps mitigate privacy concerns by avoiding tracking precise geographical coordinates. By focusing on room-level location information, we aim to balance the need for context in activity recognition while respecting user privacy. However, handling room-level location data carefully and implementing appropriate safeguards is crucial. This includes anonymizing or encrypting the data, adopting secure storage and transmission protocols, and obtaining explicit user consent. Upholding user privacy and complying with relevant data protection regulations are paramount considerations in our research. Once the application is released to the public, all the necessary security measures will be taken care of.

\section{Conclusion and Future Work}

The basic activity recognition is carried out on the WISDM dataset. The GRU model shows an accuracy of 94\%. However, once deployed on the mobile phone, the accuracy is downsized to 92\%. This is because of the testing environment and varied users while recognizing the activity. New users' data is added to the existing data, and the WISDM Fusion data is created with additional users. Adding the new user's data improved the accuracy to 93\% after deployment. Also, new user data is vital for creating long-hour data. The Two-minute, Five-minute and Ten-minute labelling are created depending on the requirement of pattern recognition. Ambient sensor data is added while creating the derived activities. By adding ambient sensors like TV On/Off status and person occupancy detection, we derived more activities than the basic activities. The more derived activities allowed us to analyze more patterns of the users. User profiling is created depending on the duration and frequency of activities. 

The work can be extended to a model for human pattern recognition using data mining techniques. Also, once the patterns are detected, user-specific profiling can be done to improve the person's behaviour. More patterns based on the sequence of the activities can be derived and analyzed. The specific models for anomaly detection can be experimented with by considering the patterns.

	\backmatter
	
	%
	%
	%
	
	%
	%

	\section{Conflict of interest}
	On behalf of all authors, the corresponding author states that there is no conflict of interest.
	%
	%
	
	\bigskip

	\bibliography{sn-bibliography}
	
	
\end{document}